\documentclass[useAMS,usenatbib]{mn2e}

\usepackage[utf8x]{inputenc}
\usepackage{bm,latexsym,tabularx,dcolumn}
\usepackage{amsmath,amsfonts,amssymb}
\usepackage{graphicx,epsfig,color,subfigure,sidecap,float}
\usepackage{longtable}
\usepackage{natbib}

\newcommand{\bea}{\begin{eqnarray}}
\newcommand{\eea}{\end{eqnarray}}
\newcommand{\be}{\begin{equation}}
\newcommand{\ee}{\end{equation}}
\newcommand{\f}{\frac}
\newcommand{\df}{\dfrac}
\newcommand{\dl}{\delta}

\newcommand{\bc}{\begin{center}}
\newcommand{\ec}{\end{center}}
\newcommand{\tm}{\times}

\newcommand{\T}{\rule{0pt}{3.6ex}}

\newcommand{\B}{\rule[-1.0ex]{0pt}{0pt}}

\title{The clustering of dark matter halos: scale-dependent bias on 
quasi-linear scales}
\begin{document}

\author[C. Jose et al.]
{Charles Jose$^{1,2}$\thanks{charlesmanimala@gmail.com},
Cedric G. Lacey$^{3}$
and Carlton M. Baugh$^{3}$ \\
$^1$SB College, Changanassery, Kottayam 686101, Kerala, India \\
$^2$IUCAA,Post Bag 4, Pune University Campus, Ganeshkhind, Pune 411007, India\\
$^3$Institute for Computational Cosmology, Department of Physics, University of Durham, South Road, Durham DH1 3LE}

\maketitle

\begin{abstract}
We investigate the spatial clustering of dark matter halos, collapsing from 
$1-4 \sigma$ fluctuations, in the redshift range $0 - 5$ using N-body 
simulations. 
The halo bias of high redshift halos ($z \geq 2$) is found to be strongly 
non-linear and scale-dependent on quasi-linear scales that are larger than 
their virial radii ($0.5-10$ Mpc/h).  
However, at lower redshifts, the scale-dependence of non-linear bias is weaker 
and and is of the order of a few percent on quasi-linear scales at $z \sim 0$.  
We find that the redshift evolution of the scale dependent bias of dark matter halos 
can be expressed as a function of four physical parameters: 
the peak height of halos, the non-linear matter correlation function 
at the scale of interest, an effective power law index of the {\it rms} 
linear density fluctuations and the matter density of 
the universe at the given redshift.  
This suggests that the scale-dependence of halo bias is not a universal 
function of the dark matter power spectrum, which is commonly assumed. 
We provide a fitting function for the scale dependent 
halo bias as a function of these four parameters. 
Our fit reproduces the simulation results to an accuracy of better 
than 4\% over the redshift range $0\leq z \leq 5$. 
We also extend our model by expressing the non-linear 
bias as a function of the linear matter correlation 
function.
It is important to incorporate our results into the clustering models  
of dark matter halos at any redshift, including those hosting early 
generations of stars and galaxies before reionization.

\end{abstract}


\section{Introduction}

The spatial distribution of luminous galaxies is a 
valuable resource for probing cosmology and the physics of galaxy formation. 
The clustering of the galaxy distribution is shaped by the clustering 
of the dark matter halos which host them.  
The clustering of dark matter halos can be quantified using the halo bias which  
describes how dark matter halos 
trace the dark matter \citep{kaiser_84,bbks_86, bond_91}. 
Conventional models assume that the halo bias is related to the 
underlying dark matter density field in a non-linear and deterministic 
fashion \citep{fry_gaztanaga_93}. 
\cite{mo_white_96} showed that, on large scales, the halo bias can be approximated 
as a scale independent function of the mass of the halos. 
In particular, they showed that the clustering of dark matter halos 
is proportional to that of the dark matter with 
the constant of proportionality being called the linear halo bias. 
The approximation for the clustering of halos using scale independent, linear bias 
is expected to be valid on scales larger than the virial radii of halos where 
dark matter halo substructure is not important. 

However, the simple picture of a scale-independent halo bias has been  
shown to be inaccurate and various non-linear and non-local processes 
result in some degree of scale dependence \citep{Matsubara1999, 
Cole2005,Seo2005,angulo_05, HUFF2007,Smith2007,angulo_08,mcdonald_09,
desjacques_09,Musso2013,aseem_13}. 
Incorporating such a scale dependence of halo bias into theoretical models 
could be crucial for interpreting the clustering of galaxies. 
While several studies have focussed on the scale dependence of the bias 
on very large scales, its scale dependence on scales larger than the 
typical virial radii of dark matter halos is equally interesting. 
These scales, corresponding to comoving length scales of 
$0.1$ to a few megaparsecs, are smaller than scales where the matter 
distribution is still linear and therefore are referred to as 
quasi-linear scales. 
The scale-dependence of halo bias on these scales arises mainly due to the  
non-linear growth of matter fluctuations \citep{Smith2007} and is difficult 
to estimate using perturbative approaches because of the non-linearity 
of matter density field \citep{reed_09}. 

There have been studies in the literature of deviations from the linear bias 
approximation on quasi-linear scales using analytic techniques 
\citep{scannapieco_barkana_02, iliev_scannapieco_03,scannapieco_thacker_05} as 
well as N-body simulations \citep{hamana_01,diaferio_03,cen_dong_04,
tinker_05,gao_white_05,angulo_08, reed_09,bosch_13}.
In particular, these studies focussed on the clustering of dark matter halos 
either in the local universe \citep[eg.][]{tinker_05} or at very high 
redshifts before the reionization of the intergalactic medium \citep{reed_09}. 
In general these studies showed that the halo bias is non-linear and scale-dependent 
on quasi-linear scales, but the scale dependence weakens on large scales. 
Specifically, \cite{reed_09} find a strong scale dependence of 
halo bias on quasi-linear scales for rare dark matter halos at  
high redshift, with the scale dependence increasing with the rarity of 
the halo.

The motivation of this paper is to study the clustering of dark 
matter halos with a specific focus on the scale dependence of halo bias on 
quasi-linear scales. 
In particular, we will focus on the redshift range $0-5$ where, 
to our knowledge, no such previous studies have been carried out. 
This will help to gauge the amplitude, scale dependence and evolution of 
the bias of dark matter halos for $0 \leq z \leq 5$ and 
hence bridge the gap between other studies which focus on 
the epochs before reionization. 
We will address this issue using N-body simulations to measure 
the dark matter and halo correlation functions in the real space. 
These measurements will be used to calibrate the nature and evolution 
of the non-linear halo bias in the redshift range $0-5$ over a range 
of length scales. 
In particular, we find that the bias of dark matter halos is 
non-linear and scale dependent on quasi-linear scales. 
Furthermore, it is not possible to express this scale dependence 
in terms of the usual parameterizations and therefore one has to 
invoke additional parameters.

The organization of this paper is as follows. 
In Section 2, we compare the halo bias of rare high redshift dark matter halos 
computed from analytic models and simulations. 
In Section 3, using simulations, we probe the scale dependence and redshift 
evolution of the non-linear bias of rare halos in the redshift range $0$ to $5$ 
and obtain a fitting function to describe these effects. 
We conclude with a brief discussion of our results and their implications in 
the final section.

\begin{figure*}
\includegraphics[trim=0cm 0.0cm 0cm 0.0cm, clip=true, width =17.5cm, height=8.7cm, angle=0]
{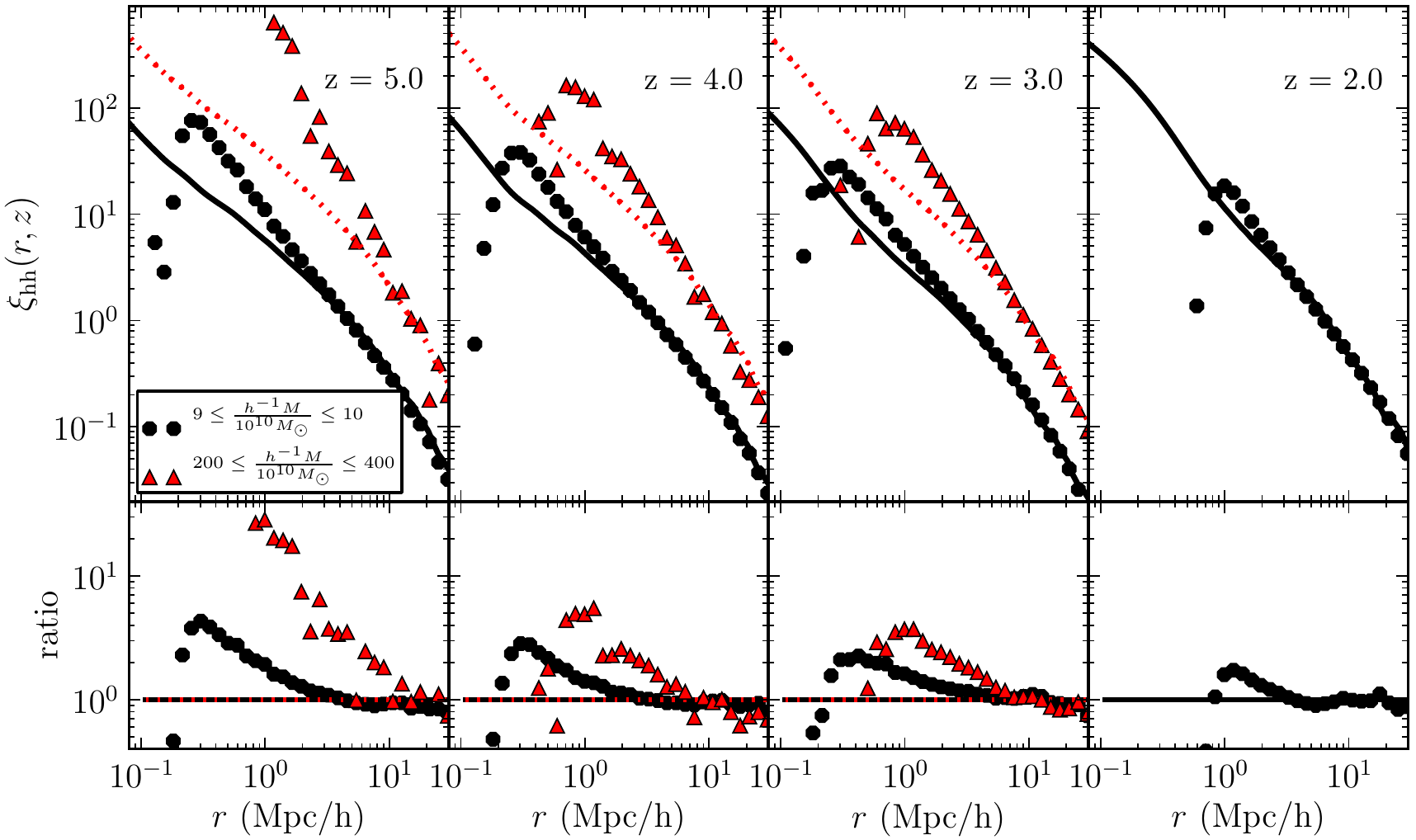} 
\caption[]
{Upper panels : The two point correlation functions of dark matter halos 
in the mass range $\sim 10^{11}-5\times 10^{12} M_\odot$ at various redshifts as labelled.
The points 
(triangles and circles) are measured from N-body simulations and the curves (solid and dotted) 
are analytic predictions using the linear bias approximation with the same 
cosmological parameters as used in the simulations. The results at $z=2$ are from 
the MXXL simulation and those at other redshifts are from the MS-W7 simulation.
Bottom panels: the ratio of the correlation functions measured from simulations to those 
computed analytically.

}
\label{fig:xir_sim_bin}
\end{figure*}

\section{Clustering of rare dark matter halos at high-z}
\label{sec:clustering}
In this section, we investigate whether the linear bias model for 
halo clustering gives a good description of the clustering of 
high-z dark matter halos on quasi-linear scales. 
For this we first describe the linear bias model for halo 
clustering in Section \ref{sec:analytic_models}. 
In Section \ref{sec:simulations} we introduce the N-body simulations 
used in our study. The clustering of dark matter halos 
estimated from these simulations is then compared with the 
clustering prediction using the linear bias model. 

\subsection{The linear bias model for halo clustering}
\label{sec:analytic_models}
In the linear bias approximation, the cross-correlation between 
halos of mass $M'$ and $M''$ is given by 
\be
\xi_{\rm hh}(r|M', M'',z) = b(M',z) b(M'',z) \xi_{\rm mm}(r,z), 
\label{eq:lin_bias_approx}
\ee
where $\xi_{\rm mm}(r,z)$ is the non-linear two point correlation function of matter 
density contrast at redshift $z$ and $b(M,z)$ is the scale independent 
linear bias of halos of mass $M$ at this redshift. 
Eq.~(\ref{eq:lin_bias_approx}) is valid on large scales, where density perturbations 
grow linearly with redshift \citep{cooray_sheth_02}. 
The two-point matter correlation function is obtained by Fourier transforming 
the non-linear matter power spectrum, $P(k,z)$ \citep{halofit03} 
\be
\xi_{\rm mm}(r,z)  = \int\limits_0^\infty \df{dk}{2\pi^2}~k^2~P(k,z)~\df{\sin(kr)}{kr}.
\label{eqn:ximm} 
\ee

It is well known that the scale independent halo bias can be expressed 
as a function of the `peak height', $\nu(M,z) = \dl_c/\sigma(M,z)$,  
of dark matter halos \citep{mo_white_96,sheth_tormen_99,sheth_tormen_02,
cooray_sheth_02,tinker_10}. 
The peak height is a measure of the rarity of  
halos \citep{sheth_tormen_02} with rarer halos having larger $\nu(M,z)$. 
Here, $\dl_c=1.686$ is the critical density for 
halo collapse and $\sigma(M,z)$ is the {\it rms} linear density fluctuation
on a mass scale $M$
\be
\sigma^2(M,z) = \sigma^2(R,z) = \int\limits_0^\infty{\frac{dk}{2\pi^2} k^2  P^{lin}(k,z) W^2(k,R)},  
\label{eqn:sigma}
\ee
where $R$ is the comoving radius of a sphere containing mass $M$, $W(k,R)$ 
is the Fourier tranformation of the top hat window function 
and $ P^{lin}(k,z)$ is the linear matter power spectrum. 

In particular, for linear halo bias, we use the fitting function 
of \citet{tinker_10} which was calibrated against N-body simulations 
and is given by, 
\be
b(M,z) = b(\nu(M,z)) = 1 - A \df{\nu^a}{\nu^a + \delta_c^a} + B \nu^b + C \nu^c. 
\label{eq:tbias}
\ee 
\citet{tinker_10} estimate the free parameters of Eq.~\ref{eq:tbias}  
to be $A = 1.0, a=0.132, B=0.183, b=1.5, C=0.265$ 
and $c=2.4$. 
The halo bias given by 
Eq.~(\ref{eq:tbias}) increases with increasing $\nu(M,z)$. 

We assume that dark matter halos which can host galaxies have a
spherical over-density $\Delta= 200$ times the average density 
of universe\footnote{\citet{tinker_10} calibrate their fitting 
function for the large scale bias as a function of $\Delta$. 
Here, the quoted parameter values are for halos with $\Delta =200$.}. 
Then, the virial radius $r_{200}$ of a halo of mass 
$M$ is $M = (4/3) \pi r^3_{200} \rho_c \Delta$. 
Under the above assumptions the halo correlation function is  
\bea
1+\xi_{\rm hh}(r|M',M'', z) = \left[1+ b(M',z) b(M'',z)  \xi_{\rm mm}(r) \right]\nonumber \\
                                    \Theta[r-r_{\rm min}(M',M'')],  
\label{eq:xi_hh}
\eea
where the function $\Theta[r-r_{\rm min}(M',M'')]$ incorporates halo exclusion to ensure that 
$\xi^{\rm hh}(r|M',M'', z) = -1 $ for $r_{\rm min} = {\rm max}[r_{200}(M'),r_{200}(M'')]$. 

The two point correlation function of dark matter halos in a mass 
bin $M' \leq M \leq  M''$ is given by 
\bea
1+\xi_{\rm hh}(r|[M',M''],z) =  \df{1}{n^2([M',M''],z)} \int\limits_{M'}^{M''} dM_1 \int\limits_{M'}^{M''} dM_2  \nonumber \\
  n(M') n(M'') \left[ 1+\xi_{\rm hh}(r|M_1,M_2, z) \right],
\label{eq:xi_2h}
\eea
where $n([M',M''],z) = \int^{M''}_{M'} dM n(M,z)$ is the total number density of 
halos in the mass bin [$M', M''$]. For the halo mass function, $n(M,z)$, we use the 
fitting function of \cite{jenkins_01} which is in excellent agreement 
with the mass functions obtained from the simulations used in our study. 
Eq.~(\ref{eq:xi_2h}) is a reasonable approximation for the usual 2-halo term 
for halo clustering on scales larger than the virial radii of dark matter 
halos \citep{cooray_sheth_02}. 

The average bias of halos with mass between $M'$ and $M''$, on 
scales bigger than their virial radii can be written as being scale-independent 
and is given by  
\be
b([M',M''],z) = \df{1}{n([M', M''],z)} \int\limits^{M''}_{M'} dM b(M,z) n(M,z). 
\ee
Using this result for the bias in Eq.~(\ref{eq:xi_2h}), we get 
\be
\xi_{\rm hh}(r|[M',M''],z)  =  b^2([M',M''],z) \xi_{\rm mm}(r,z).
\label{eq:xi_approx}
\ee
In what follows, we will compute the halo correlation functions 
for dark matter halos in mass bins and compare with those measured from 
N-body simulations. 

\subsection{N -body simulations}
\label{sec:simulations}
Our study mainly uses two cosmological dark matter N-body 
simulations, the MS-W7 simulation \citep{guo_12,pike_14} and the 
Millennium-XXL or MXXL simulation \citep{mxxl_12}. 
The MS-W7 simulation uses a cubic computational box of comoving length 
500 $h^{-1}$ Mpc with 2160$^3$ particles of mass $8.61 \times 10^8 M_\odot$.  
This is used to probe the clustering of halos at $z=3,4$ and $5$. 
It adopts a flat $\Lambda$CDM background cosmology, 
which is in agreement with the WMAP7 results \citep{larson_11_wmap7}, with $ h= 0.704$, 
$\Omega_b=0.0455$, $\Omega_c= 0.2265$, $\Omega_\nu = 0.0$, 
$\sigma_8=0.81$ and $n_s= 0.967$.

The MXXL extends the previous Millennium and Millennium-II simulations 
\citep{springel_white_05, millenium2_09} and follows the evolution of 
$6720^3$ dark matter particles inside a cubic box of length 
3000 $h^{-1}$ Mpc. The particle mass is $8.46 \times 10^9 M_\odot$.  
This simulation adopts a $\Lambda$CDM cosmology with the
same cosmological parameters as the previous Millennium simulations. 
Accordingly, $ h= 0.73$, $\Omega_b=0.045$, $\Omega_c= 0.205$, 
$\Omega_\nu = 0.0$, $\sigma_8=0.9$ and $n_s= 1.0$.  
The MXXL halos are used to investigate the clustering at $z=0,1,2$ 
and $3$. We also compare the results obtained using MXXL 
simulation at $z=3$ with those obtained using using the Millennium simulation 
at the same redshift, which has a box of 500 $h^{-1}$ Mpc.

In both simulations, groups of more than 20 particles are identified as dark 
matter halos using a friends-of-friends algorithm (FOF(0.2)) with linking parameter 
equal to $0.2$ of the mean particle separation \citep{davis_1985}.  
The halo mass functions from these simulations are well described by 
the fitting function of \cite{jenkins_01} over a wide range of halo masses 
and to an accuracy better than 10 \%. 

The two point correlation functions of dark matter halos and dark matter 
particles from the simulations are computed by counting the number of pairs 
as a function of the separation, $r$, relative to that of a random 
distribution and is given by 

\be
\xi^{\rm sim}(r) = \df{N^{p}(r)}{N^{p}_{ran}(r)} -1 
\label{eq:xifromsimul}
\ee
where $N_{p}(r)$ is the total number of pairs in the simulation separated 
by a distance $r$ to $r+\delta r$ and $N^{p}_{ran}(r)$ is the total 
number of pairs. 

As mentioned above, in this paper, we focus on the clustering of rare dark 
matter halos on quasi-linear scales in the redshift range $0-5$. 
Therefore, we consider only those halos with a peak 
height $\nu(M,z) > 1$. At $z \sim 0$, the typical 
masses of these halos range between $10^{13} - 10^{15} M_\odot$ 
and therefore they correspond to poor galaxy groups and clusters. 
On the other hand, for $z \geq 2$, the masses of rare halos range from 
$10^{10}-10^{13} M_\odot$. 
As we see later, the scale dependence of the halo bias due to non-linear 
clustering is much more significant at higher redshifts ($z = 2-5$) 
than it is at lower redshifts. 
Therefore, we first address the issue of the non-linear clustering of 
high redshift dark matter halos on quasi-linear scales and then its evolution in the 
low redshift universe.

\subsection{Comparing simulations and linear bias models}

We first show that the clustering strength of high-z, rare dark matter 
halos on quasi-linear scales differs significantly from the predictions 
of the linear bias model by comparing with the spatial correlation functions 
estimated from N-body simulations. 
In the top panels of Fig.~\ref{fig:xir_sim_bin}, the halo correlation 
functions estimated from simulations ($\xi^{\rm sim}_{\rm hh}(M,r,z)$) 
are shown at $z=2,3,4$ and $5$ 
for halos in the mass range $9 \times 10^{10}-10^{11}$ $h^{-1} M_\odot$ (black circles) and 
$2 \times 10^{12}-4 \times 10^{12}$ $h^{-1} M_\odot$ (red triangles).
We note that, these halos respectively host typical Lyman-$\alpha$ emitters (LAEs) and 
Lyman-break galaxies (LBGs) in the same redshift 
range \citep{charles_12_clustering,charles_13_lae_clustering} and 
are rare halos, collapsing from $2-3 \sigma$ fluctuations ($\nu(M,z) \sim 2-3$). 
On small scales the correlation functions drop to $-1$ due to 
halo exclusion. These scales correspond to the typical virial radius of halos 
in the given mass bin. 

Also shown in the top panels of Fig.~\ref{fig:xir_sim_bin} are the correlation 
functions, $\xi_{\rm hh}(M,r,z)$, predicted by 
Eq.~(\ref{eq:xi_approx}) for the same cosmological parameters as used in the 
simulations. 
The correlation functions are computed for halos in the same mass 
bins used to estimate $\xi^{\rm sim}_{\rm hh}(M,r,z)$. 
Fig.~\ref{fig:xir_sim_bin} clearly shows that $\xi^{\rm sim}_{\rm hh}(r,z)$ and $\xi_{\rm hh}(r,z)$ agree 
well with each other on large scales ($r \gtrsim 10-15$ $h^{-1}$ Mpc).  
However, on quasi-linear scales ($r \sim 0.5-10$ $h^{-1}$ Mpc), 
$\xi^{\rm sim}_{\rm hh}(r,z)$ determined from the simulations shows an  
excess compared to $\xi_{\rm hh}(r,z)$ computed using Eq.~(\ref{eq:xi_approx}).  

To understand the degree of this deviation more clearly, we have plotted 
in the bottom panels of Fig.~\ref{fig:xir_sim_bin} the ratio of the dark 
matter halo correlation functions measured from simulations to that computed from 
the linear bias model (i.e $\xi^{\rm sim}_{\rm hh}(r,z)/\xi_{\rm hh}(r,z)$) for each mass bin. 
It is clear from the figure that, on quasi-linear scales, the predictions of 
the scale-independent bias model are insufficient to explain the halo 
correlation functions measured directly from the simulations. 
For example, the massive halos at the highest redshift 
($2\times 10^{12} \leq M/M_\odot \leq 4\times 10^{12}$ at $z=5$) 
show clustering in the simulations that is sometimes larger by a factor as large as 
$\sim$ 20 at $ 1 \leq  r \leq 2$ Mpc/h, compared to the linear bias model predictions. 
On the other hand, at lower redshifts and for less massive halos 
($9\times 10^{10} \leq M/M_\odot \leq 10^{11}$ at $z=3$), the clustering excess  
is only a factor of $2-3$ at $r \sim 0.5$ Mpc/h. 
Furthermore, the deviation between $\xi^{\rm sim}_{\rm hh}(r,z)$ and $\xi_{\rm hh}(r,z)$ increases 
with the redshift and mass of dark matter halos. 
Overall we conclude that the halo bias of high redshift dark matter halos is strongly 
scale dependent on quasi-linear scales and the scale dependence increases with 
the rarity of the halos. 

Earlier studies focused on the non-linear bias of halos at the present epoch 
\citep{hamana_01,diaferio_03,tinker_05} or at redshifts before 
reionization \citep{reed_09}. The scale dependence of the non-linear bias in the 
fitting functions provided by \cite{hamana_01,diaferio_03,tinker_05} is
too weak to explain the clustering of halos at the redshifts, masses and scales of 
interest here. 
The fitting function of \cite{reed_09} has a stronger scale-dependence 
and describes the non-linear clustering of high redshift MS-W7 halos 
correctly. However, as we see 
later, their results are not consistent with the bias measured from the MXXL simulation 
and also at lower redshifts ($z=0-2$). 
Therefore, non-linear clustering of rare halos on quasi-linear scales 
has not been satisfactorily addressed and thus warrants 
further investigation. 

\begin{figure}
\includegraphics[trim=0cm 0.0cm 0cm 0.0cm, clip=true, width =8.0cm, height=8.0cm, angle=0]
{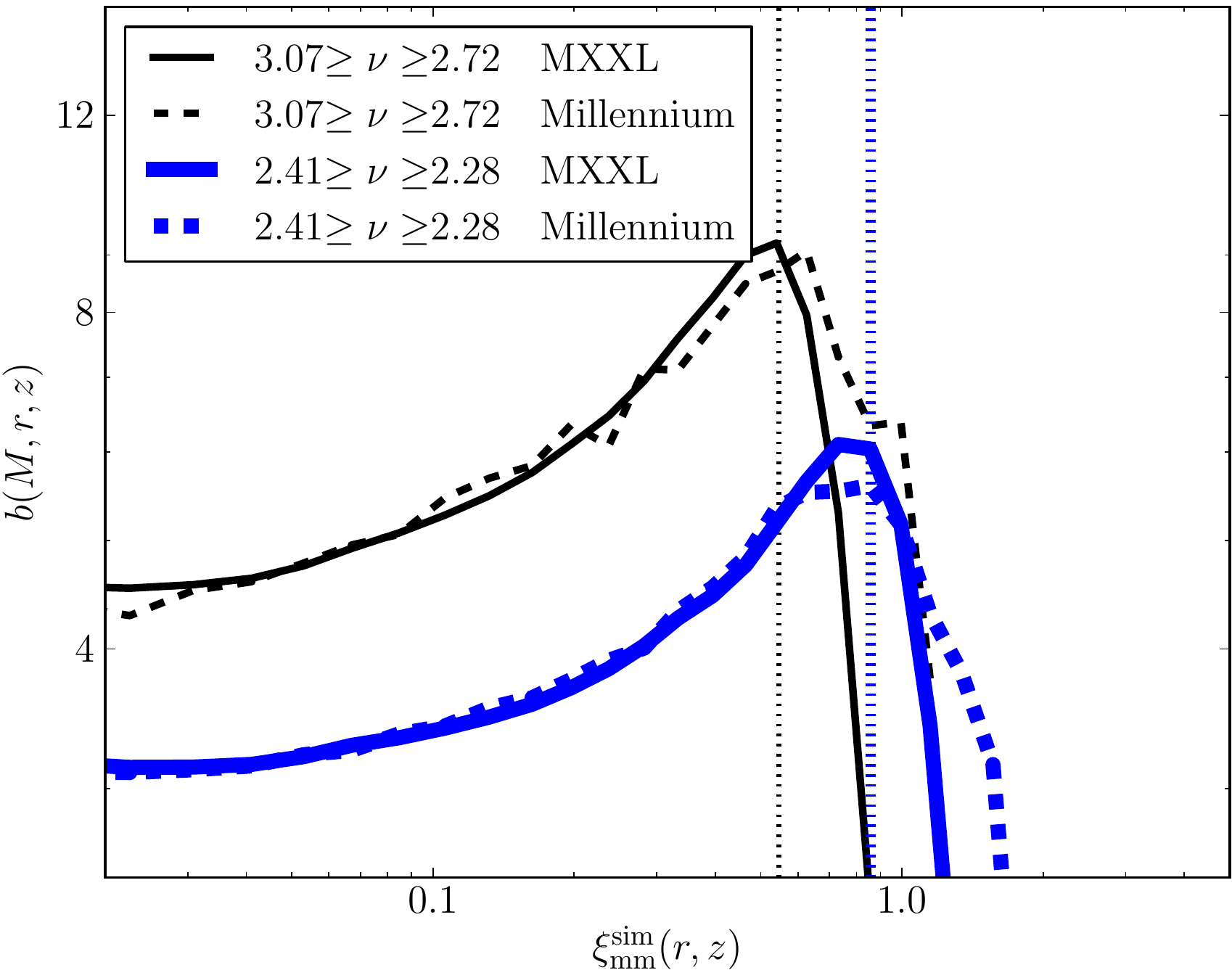} 
\caption[]
{
The halo bias $b(r,M,z) = \sqrt{\xi^{\rm sim}_{\rm hh}(r,z)/\xi^{\rm sim}_{\rm mm}(r,z)}$ computed 
at $z=3$ using Millennium and MXXL halos in $\nu(M,z)$ bins, as indicated 
by the label. The thick (blue) and thin (black) lines respectively correspond 
to halos with $2.41 \geq \nu \geq 2.28$ and $3.07 \geq \nu \geq 2.72$. 
The thick (blue) and thin (black) vertical lines indicate scales corresponding 
to twice the virial radius of the most massive halo in each sample. 
}
\label{fig:b_mxxl_mi}
\end{figure}

\begin{figure*}
\includegraphics[trim=0cm 0.0cm 0cm 0.0cm, clip=true, width =14.0cm, height=14cm, angle=0]
{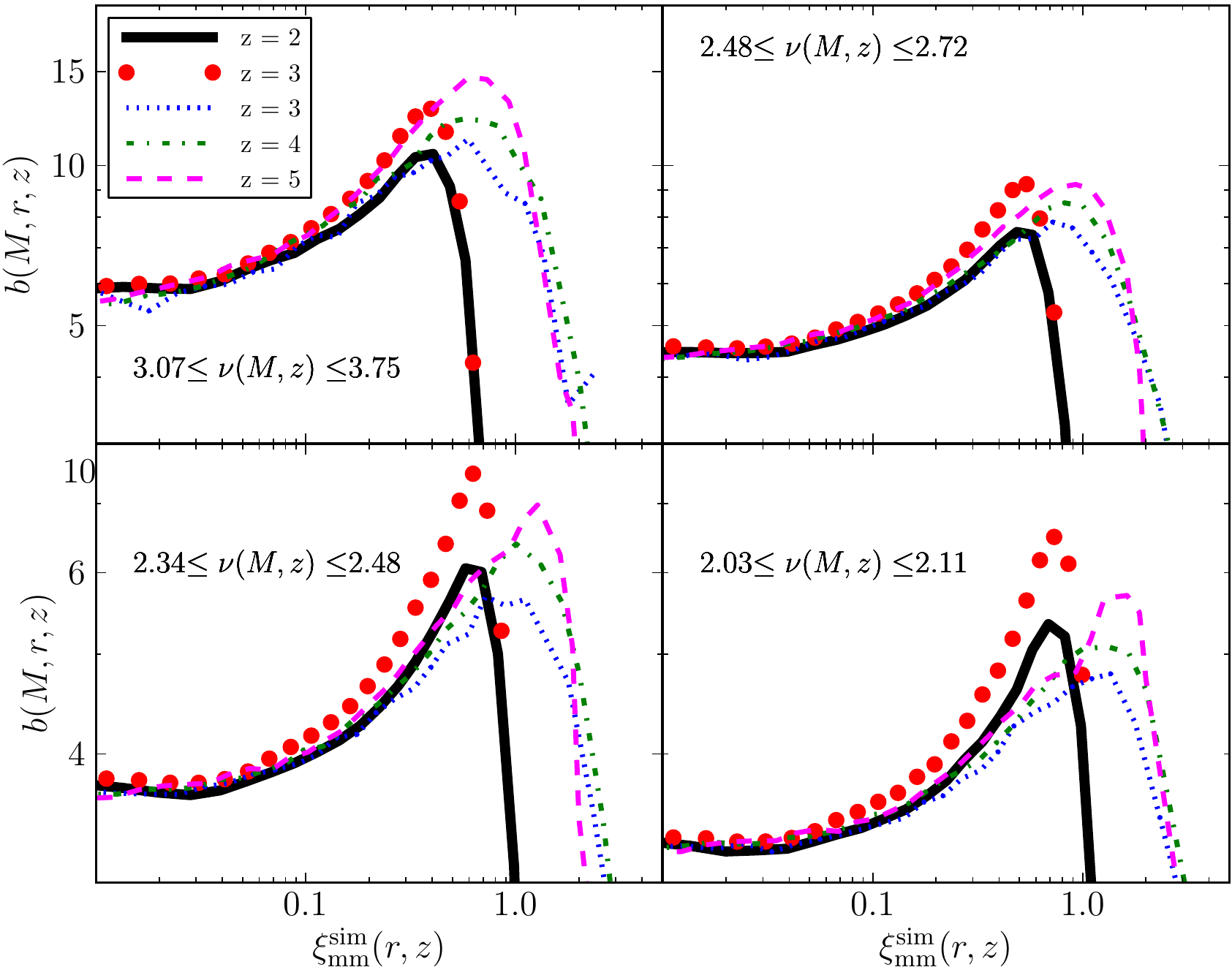} 
\caption[]
{The halo bias $b(r,M,z) = \sqrt{\xi^{\rm sim}_{\rm hh}(r,z)/\xi^{\rm sim}_{\rm mm}(r,z)}$ plotted as a function 
of $\xi^{\rm sim}_{\rm mm}(r,z)$ in the redshift range $2-5$. Each panel shows the $b(r,M,z)$ 
of halos in $\nu(M,z)$ bins, as indicated by the label. 
The results obtained from the MXXL simulation are shown by the solid black lines ($z=2$)
and red circles ($z=3$) whereas other curves corresponds to the results from the 
MS-W7 simulation. 
}
\label{fig:br_universal_fn}
\end{figure*}

\begin{table*}
\tabcolsep 3.0pt
\begin{tabular}{|cc|cccc|ccc|c|}
\hline
\multicolumn{2}{|c|}{}   &\multicolumn{8}{|c|}{$M_{av}/M_\odot$} \B \B \B   \T\T \\
\cline{3-10}
$(\nu_{\rm min},\nu_{\rm max})$ &$\nu_{av}$    &\multicolumn{4}{|c|}{MXXL halos}   &\multicolumn{3}{|c|}{MS-W7 halos} &\multicolumn{1}{|c|}{Millennium halos}    \B \B \B   \T\T \\
\cline{3-10}
~  &  &z=0  &z=1  &z=2 &z=3  &z=3  &z=4  &z=5  &z=3  \B \B \B \B \T \T  \T\T \\
\hline 
\T \B ($3.37,4.21$) &$3.77$ &$1.9\tm 10^{15}$ &$3.0\tm 10^{14}$ &$6.1\tm 10^{13}$ &$1.5\tm 10^{13}$ &$8.0\tm 10^{12}$ &$2.7\tm 10^{12}$ &$7.4\tm 10^{11}$ &$1.5\tm 10^{13}$\\
\T \B ($3.07,3.37$) &$3.22$ &$1.2\tm 10^{15}$ &$1.8\tm 10^{14}$ &$3.3\tm 10^{13}$ &$7.7\tm 10^{12}$ &$3.9\tm 10^{12}$ &$1.2\tm 10^{12}$ &$3.1\tm\T \B  10^{11}$ &$7.7\tm 10^{12}$ \\
\T \B ($2.81,3.07$) &$2.93$ &$8.9\tm 10^{14}$ &$1.2\tm 10^{14}$ &$2.1\tm 10^{13}$ &$4.5\tm 10^{12}$ &$2.2\tm 10^{12}$ &$6.5\tm 10^{11}$ &$1.6\tm 10^{11}$ &$4.5\tm 10^{12}$ \\
\T \B ($2.59,2.81$) &$2.70$ &$6.5\tm 10^{14}$ &$8.2\tm 10^{13}$ &$1.3\tm 10^{13}$ &$2.7\tm 10^{12}$ &$1.3\tm 10^{12}$ &$3.6\tm 10^{11}$ &$8.0\tm 10^{10}$ &$2.7\tm 10^{12}$\\
\T \B ($2.44,2.59$) &$2.52$ &$5.0\tm 10^{14}$ &$5.9\tm 10^{13}$ &$9.2\tm 10^{12}$ &$1.8\tm 10^{12}$ &$8.0\tm 10^{11}$ &$2.1\tm 10^{11}$   &  &$1.8\tm 10^{12}$\\
\T \B ($2.28,2.41$) &$2.34$ &$3.7\tm 10^{14}$ &$4.2\tm 10^{13}$ &$6.0\tm 10^{12}$ &$1.1\tm 10^{12}$ &$4.8\tm 10^{11}$ &$1.2\tm 10^{11}$   &  &$1.1\tm 10^{12}$\\
\T \B ($2.19,2.28$) &$2.23$ &$3.0\tm 10^{14}$ &$3.3\tm 10^{13}$ &$4.6\tm 10^{12}$   &&$3.4\tm 10^{11}$ &$8.2\tm 10^{10}$   &   &$8.0\tm 10^{11}$\\
\T \B ($2.11,2.19$) &$2.15$ &$2.6\tm 10^{14}$ &$2.7\tm 10^{13}$ &$3.6\tm 10^{12}$   &&$2.5\tm 10^{11}$ &$5.9\tm 10^{10}$   &   &$6.1\tm 10^{11}$\\
\T \B ($2.03,2.11$) &$2.07$ &$2.2\tm 10^{14}$ &$2.2\tm 10^{13}$ &$2.9\tm 10^{12}$   &&$1.9\tm 10^{11}$   &  &   &$4.7\tm 10^{11}$\\
\T \B ($1.96,2.03$) &$2.00$ &$1.9\tm 10^{14}$ &$1.8\tm 10^{13}$ &$2.3\tm 10^{12}$   &&$1.4\tm 10^{11}$   &  &   &$3.6\tm 10^{11}$\\
\T \B ($1.87,1.92$) &$1.89$ &$1.5\tm 10^{14}$ &$1.4\tm 10^{13}$ &$1.6\tm 10^{12}$   &&$9.2\tm 10^{10}$   &  &   &$2.5\tm 10^{11}$\\
\T \B ($1.77,1.81$) &$1.79$ &$1.2\tm 10^{14}$ &$1.0\tm 10^{13}$ &$1.1\tm 10^{12}$   &&$5.8\tm 10^{10}$   &  &   &$1.6\tm 10^{11}$\\
\T \B ($1.69,1.72$) &$1.70$ &$9.2\tm 10^{13}$ &$7.5\tm 10^{12}$   &  &  &  &  &    &$1.1\tm 10^{11}$\\
\T \B ($1.53,1.56$) &$1.55$ &$5.8\tm 10^{13}$ &$4.2\tm 10^{12}$   &  &  &  &  &    &$4.9\tm 10^{10}$\\
\T \B ($1.30,1.32$) &$1.31$ &$2.5\tm 10^{13}$ &$1.5\tm 10^{12}$   &  &  &  &  & &\\
\T \B ($1.12,1.14$) &$1.13$ &$1.1\tm 10^{13}$   &  &  &  &  &  &     & \\
\hline
\end{tabular}
\caption[]{
Column 1: The $\nu(M,z)$ bins of halos used to calibrate the 
non-linear bias. Column 2: The corresponding average peak height, $\nu_{av}$. 
Other columns give the average mass, $M_{av}$, of halos in the sample at the 
given redshift. If $M_{av}$ is not given, then the $\nu(M,z)$ bin 
is not used in our analysis. Columns 3-6 results are for MXXL halos, 
columns 7-9 are for MS-W7 halos whereas column 10 refers to Millennium 
halos.  
}
\label{tab1}
\end{table*}

\section{The scale-dependent, non-linear halo bias}
\subsection{The measured bias}
We have shown in the previous section that, on quasi-linear scales, high-z dark 
matter halos collapsing from $2-3$ $\sigma$ fluctuations cluster more strongly than 
the predictions of the linear bias model. Therefore, to understand the 
clustering of these rare halos, one has to invoke a scale-dependent, non-linear bias.  
For this, we first define a non-linear, scale dependent halo bias of dark 
matter halos at any redshift as \citep{scannapieco_barkana_02, reed_09} 
\be
b_{nl}(r,M,z) = \sqrt{\df{\xi^{\rm sim}_{\rm hh}(r,z)}{\xi^{\rm sim}_{\rm mm}(r,z)}}.  
\label{eq:br}
\ee
Here, $\xi^{\rm sim}_{\rm mm}(r,z)$ is the non-linear dark matter correlation function
computed directly from the simulations using Eq.~(\ref{eq:xifromsimul}). 
The function $b(r,M,z)$ is thus expected to be independent of $r$ on large scales.  

We note that there are alternative definitions of the halo bias in 
Fourier space and also as the ratio of the halo-matter cross correlation 
function to the matter correlation function (eg. \citealt{tinker_10,
manera11}). 
The choice of a particular definition of the halo bias can in principle introduce a 
scale-dependence in the bias measured from the simulations \citep{baumann_13}. 
This scale-dependence introduces a few percent difference 
between the measured bias that is defined by Eq.~(\ref{eq:br}) and 
those defined by 
other bias definitions \citep{smith_marian_11, pollack_12}. 
This effect is much weaker compared to the strong scale-dependence of the halo bias 
presented in this work and hence will be neglected. 
We also note that $b_{nl}(r,M,z)$, defined by Eq.~(\ref{eq:br}), encapsulates the scale-dependence of 
the halo bias due to the non-linear higher order correlations of the matter 
distribution and also due to the scale-dependence of the linear halo 
bias. 
Thus, one can directly use $b_{nl}(r,M,z)$ to compute the $\xi_{hh}$ at any given  
scale in real space using Eq.~(\ref{eq:xi_hh}) with minimum ambiguities, which 
is our primary goal.

In the previous section, we estimated the clustering of halos in different mass bins. 
However, it would be more useful if the non-linear bias could be calculated 
from the dark matter power spectrum. One could then hope to apply 
our results in more general contexts. Furthermore, as discussed in the previous 
section, we first focus on high redshift halos.   
Therefore, we measured the scale-dependent halo bias ($b_{nl}(r,M,z)$) of dark 
matter halos from the MS-W7, MXXL and Millennium simulations in 
bins of peak height, $\nu(M,z)$, in the 
redshift range $2-5$.

First, we briefly discuss the effects of resolution and the halo exclusion 
effect on the estimated non-linear bias by comparing results obtained 
using MXXL and Millennium halos at $z=3$.  
As discussed before, these simulations have the same set of cosmological 
parameters, but different mass resolutions and volumes. 
In Fig.~\ref{fig:b_mxxl_mi}, we have plotted the non-linear and 
scale-dependent bias at $z=3$ 
for halos (in a given $\nu$ bin) from the MXXL and Millennium simulations. 
Also shown by the vertical line, is the length scale corresponding to twice the 
virial radius of the most massive halo in each sample. 

We first note that in Fig.~\ref{fig:b_mxxl_mi}, the scale-dependent bias 
at $z=3$ is expressed as a function of $\xi^{\rm sim}_{\rm mm}(r,z)$ 
at the same redshift. This choice has been made in several analytic and numerical 
studies probing the clustering of dark matter halos on quasi-linear scales 
\citep{hamana_01,scannapieco_barkana_02, diaferio_03, tinker_05, reed_09,bosch_13}. 
These studies present the scale-dependent bias, $b_{nl}(r,M,z)$ as a 
function of the non-linear dark matter correlation function, 
$\xi^{\rm sim}_{\rm mm}(r,z)$ at the same redshift. In what follows, we adopt 
this approach and express $b_{nl}(r,M,z)$ as a function of 
$\xi^{\rm sim}_{\rm mm}(r,z)$. In this approach, the scale-dependent bias can be 
thought of as a function of the non-linear dark matter power spectrum, since the 
matter correlation function is the Fourier transform of the non-linear dark 
matter power spectrum.

It is clear from Fig.~\ref{fig:b_mxxl_mi} that the halo bias estimated from 
the two simulations agree well with one another on scales larger than twice 
the virial radius of the most massive halo in the sample. However, on smaller 
scales, the estimated halo bias is different between the two simulations. 
Since both simulations use the same set of cosmological parameters, 
this could be due to the difference in mass resolution between the 
simulations.   
We also note that, on the largest scales ($\xi^{\rm sim}_{\rm mm}(r,z) < 0.05$), the bias 
is approximately a constant. On smaller scales, the non-linear bias increases 
with decreasing scale (increasing $\xi^{\rm sim}_{\rm mm}$) 
and reaches a maximum value around the scale 
corresponding to twice the virial radius of the most massive halo in the sample. 
On smaller scales than this, the halo bias drops to $0$. This suggest 
that, while probing the clustering of a sample of halos, the halo exclusion 
effect is important on scales smaller than twice the virial radius of the 
most massive halo in the sample. 
Because of these effects due to the resolution and halo exclusion, 
our further discussion and analysis will consider the clustering 
of a sample of halos only on scales larger than twice the virial radius 
of the most massive halo in that sample.

We now present our estimates of the non-linear bias from different 
simulations at various redshifts in 
Fig. \ref{fig:br_universal_fn} as a function of $\xi^{\rm sim}_{\rm mm}(r,z)$. 
The results measured from the MXXL simulation at $z=2$ are shown by solid black 
lines and at $z=3$ are shown using red circles. All the other curves 
at $z=3,4$ and $5$ are estimates 
of $b_{nl}(r,M,z) $ from the MS-W7 simulation.   
Each panel corresponds to a different bin of $\nu(M,z)$ (see labels). 
We again emphasize that $b_{nl}(M,r,z)$ at a given redshift is plotted 
against $\xi^{\rm sim}_{\rm mm}(r,z)$ at the same redshift.

We first note that, on scales corresponding to 
$\xi^{\rm sim}_{\rm mm} \lesssim 0.1$, 
$b_{nl}(r,M,z)$ measured from the MS-W7 and MXXL simulations 
agree well with each other. On smaller scales, the estimated bias is 
different for the two simulations. The halo bias measured from 
the MXXL simulation drops to zero on larger scales compared to the 
bias estimated from the MS-W7 simulation. This is because, for halos 
in a given $\nu$ bin, the masses and virial radii of MXXL halos 
are larger than those of MS-W7 halos.  
 
\subsection{A model for the halo bias} 
\label{sec:nlbias}
As discussed before, $b_{nl}(r,M,z)$ at 
different redshifts is fairly constant for $\xi^{\rm sim}_{\rm mm} \lesssim 0.05$. 
These scales typically correspond to comoving length scales greater 
than 10 $h^{-1}$ Mpc. Thus on such large scales, the expression for 
the non-linear bias reverts back to the usual scale-independent 
large scale bias, which is only a function of the peak height $\nu$ alone. 
Therefore, one can write      
\be
b_{nl}(r,M,z) = \gamma(r,M,z) b(\nu), 
\label{eq:gammadef}
\ee
where $ b(\nu)$ is the large scale bias. Here the non-linear bias, 
$b_{nl}(r,M,z) $, is written as the product of a scale-dependent 
function, $\gamma(r,M,z)$, and the large scale bias.  
The scale-dependent function $\gamma(r,M,z)$ is thus expected to be close to unity 
on large scales. 

To understand the evolution of the non-linear bias of rare halos with 
redshift, one has to calibrate the expressions for the 
large scale linear bias $b(\nu)$ and scale dependent function $\gamma(r,M,z)$. 
In what follows, we first obtain a fitting function for 
$b(\nu)$ and then constrain the functional form of $\gamma(r,M,z)$. 

\begin{figure}
\includegraphics[trim=0cm 0.0cm 0cm 0.0cm, clip=true, width =8.0cm, height=8.0cm, angle=0]
{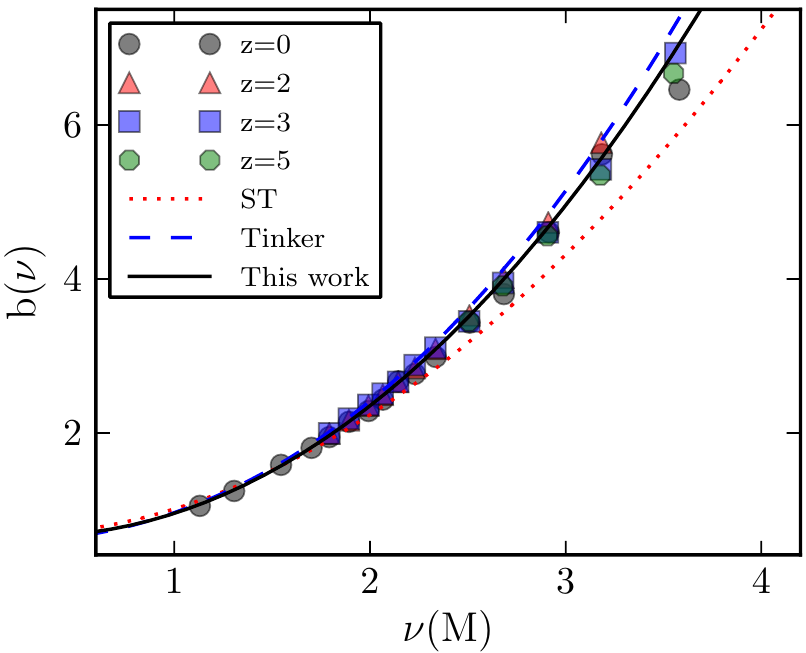} 
\caption[]
{Our fit to Eq.~(\ref{eq:tbias}) for the large scale linear halo bias (solid black curve) 
along with the simulation measurements (symbols) at various redshifts. 
The points for $z = 0 $ and $2$ are measured from the MXXL simulation and 
those at $z=3$ and $5$ are obtained from the MS-W7 simulation. 
The red dotted and blue dashed curves are the fitting functions 
of the linear halo bias given by \cite{sheth_tormen_99} (ST) 
and \cite{tinker_10} (Tinker) respectively. 
}
\label{fig:b_av}
\end{figure}

\subsection{The large scale bias}
To estimate $b(\nu)$, we measured the correlation functions of dark 
matter halos in different $\nu(M,z)$ bins in the redshift range $0-5$. 
These $\nu$-bins are given in column 1 of Table~\ref{tab1}. 
Also given in the table are the average peak height, $\nu_{av}$ and 
average mass of halos, $M_{av}$, in these bins at each redshift. 
The average peak height is given by $\nu_{av} = \delta_c / \sigma_{av}$, 
where $\sigma_{av}$ is computed as  
\be
\sigma^2_{av}(z) = \df{\int dM~ n(M,z)~\sigma^2(M,z)}{\int dM~n(M,z)}. 
\ee

We define the large scale bias, $b(\nu)$, as the average bias of halos 
that are separated by $10 \leq r \leq 25$ Mpc/h, i.e.
\be
b(\nu) = \sqrt{\df{\int^{25}_{10} dr~r^2~\xi_{\rm hh}(\nu, \xi^{\rm sim}_{\rm mm}(r))}{\int^{25}_{10} dr~r^2~\xi^{\rm sim}_{\rm mm}(r) }}. 
\label{eq:bav}
\ee
We then obtain a fitting function for $b(\nu)$ by refitting the 
free parameters of Eq.~(\ref{eq:tbias}) to the $b(\nu)$ measured from the simulations 
using Eq.~(\ref{eq:bav}).
The best fit parameters are estimated to be 
$A=1.0, a= 0.36 , B=-1.156, b= 2.18, C= -0.749$ and $c= 2.18$, treating all bias 
values with equal weight. 

In Fig.~\ref{fig:b_av}, our fit (black solid line) for $b(\nu)$ is 
overplotted with the symbols measured directly from the 
simulation at different redshifts. 
The data points at $z=0,1,$ and $2$ are measured from the MXXL simulation and 
those at $z=3,4,$ and $5$ are obtained from MS-W7 simulation. 
It is clear from Fig.~\ref{fig:b_av} that our fitting function for $b(\nu)$ agrees 
very well with the measurements from the simulations. 
In fact, we find that, the overall agreement between the 
simulation results and the fitting function is within 3\%.   
Also shown in Fig.~\ref{fig:b_av} are 
the fitting functions for halo bias from \cite{sheth_tormen_99} (ST, red dotted curve) 
and \cite{tinker_10} (Tinker, blue dashed curve). One can see from the figure 
that when $\nu(M) \lesssim 2$, our formula compares well with 
these two fitting functions (particularly with the Tinker formula). 
However, for larger values of $\nu(M)$, the ST formula 
predicts lower bias values and Tinker formula gives slightly 
larger bias values compared with our formula for halo bias. 
Thus for rarer halos with $\nu \geq 2$, our analysis predicts a slightly lower value 
for the large scale bias compared to the Tinker formula. 
However, we also note that \cite{tinker_10} use the spherical overdensity (SO) 
algorithm \citep{tinker_kravtsov_08} to identify halos, which 
is different from the FOF(0.2) algorithm used in this work. 
Such a difference in the bias of halos identified by these two algorithms 
has already been noted by \cite{tinker_kravtsov_08,tinker_10}.   
Further, the simulations used by \cite{tinker_10} span 
a wider range of cosmological parameters than used in this work. 
This could also account for the difference between the estimated 
large scale halo bias.

\subsection{The scale dependence of halo bias}

\begin{figure}
\includegraphics[trim=0cm 0cm 0cm 0cm, clip=true, width =8.0cm, height=8.0cm, angle=0]
{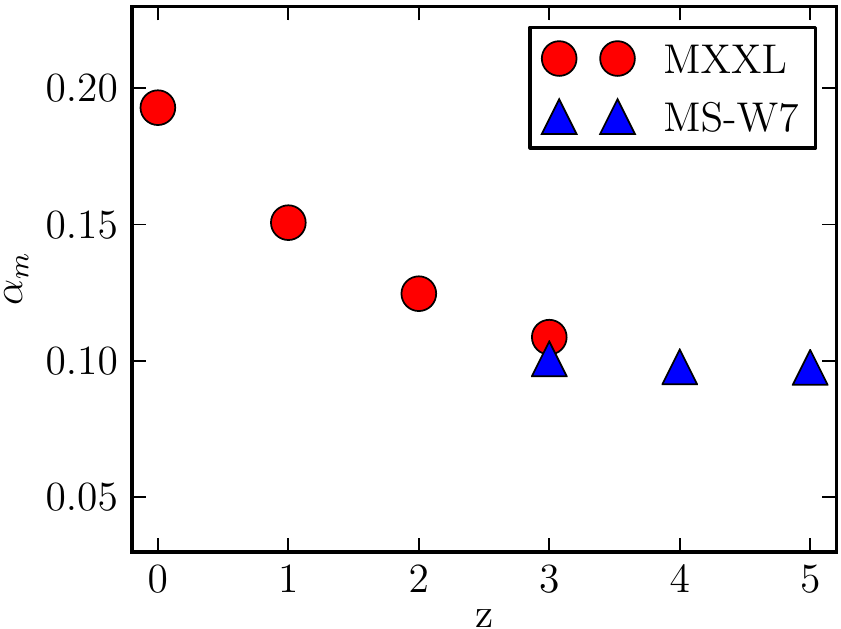} 
\caption{
The effective power law index, $\alpha_m$, (Eq.~(\ref{eqn:alpha})) is plotted as a 
function of redshift. The blue triangles and red circles correspond 
to the values obtained for the MS-W7 and the MXXL cosmology 
respectively.  
}
\label{fig:mnltomcoll}
\end{figure}

\begin{figure*}
\includegraphics[trim=0cm 0cm 0cm 0cm, clip=true, width =15.0cm, height=15.0cm, angle=0]
{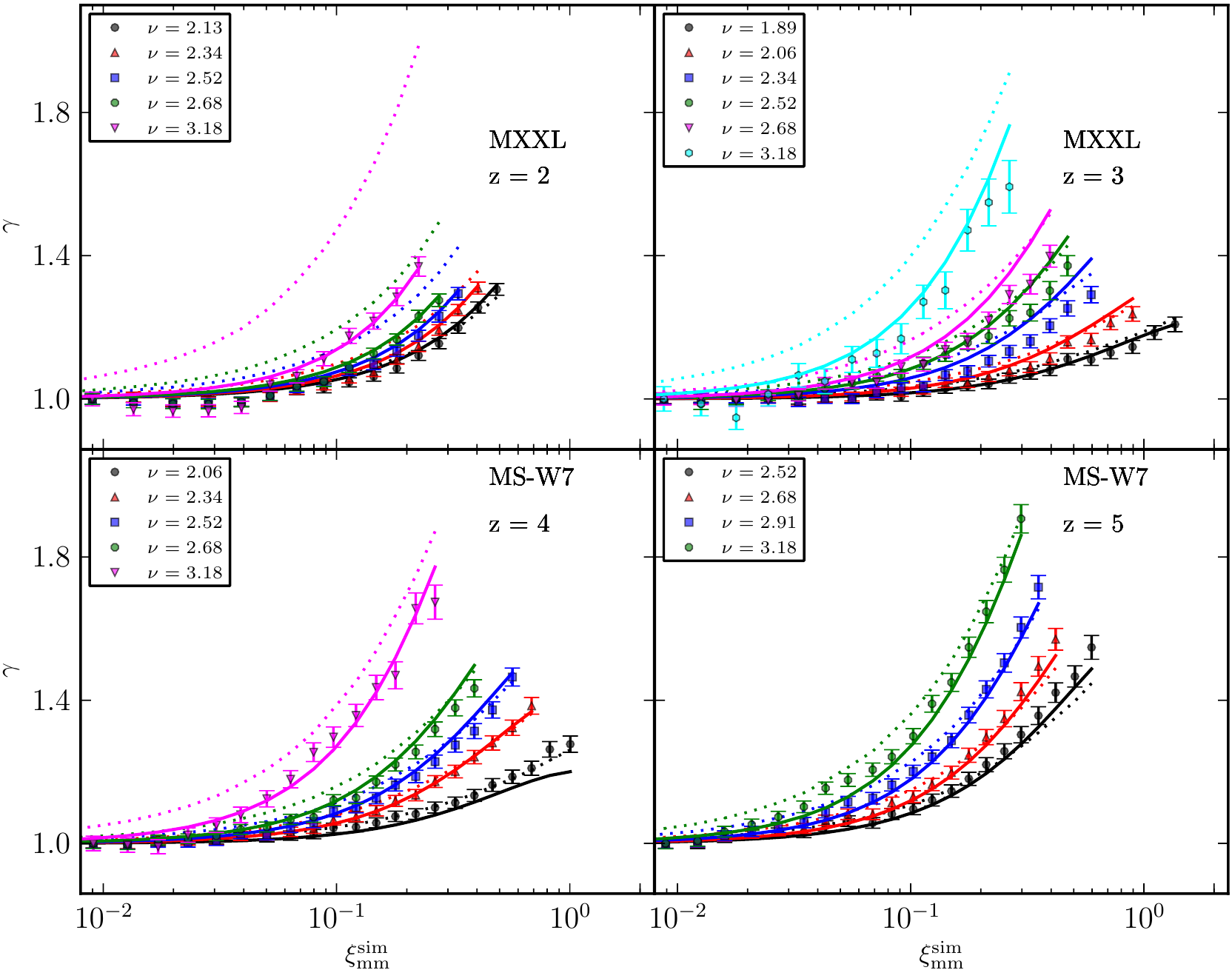} 
\caption[]
{
The scale dependence of non-linear bias, $\gamma(\xi^{\rm sim}_{\rm mm},\nu, \alpha_m)$, 
measured from N-body simulations, in the redshift range $z=2-5$ as a function of $\xi^{\rm sim}_{\rm mm}(r)$ 
for halos in the $\nu$-bins listed in the legend. 
Solid lines: the fit for $\gamma(\xi^{\rm sim}_{\rm mm},\nu, \alpha_m)$ presented in  this work. 
Dotted lines: the fitting function for $\gamma$ given by \cite{reed_09}. 
}
\label{fig:gamma_z25}
\end{figure*}

\begin{figure*}
\includegraphics[trim=0cm 0cm 0cm 0cm, clip=true, width =17.0cm, height=17.0cm, angle=0]
{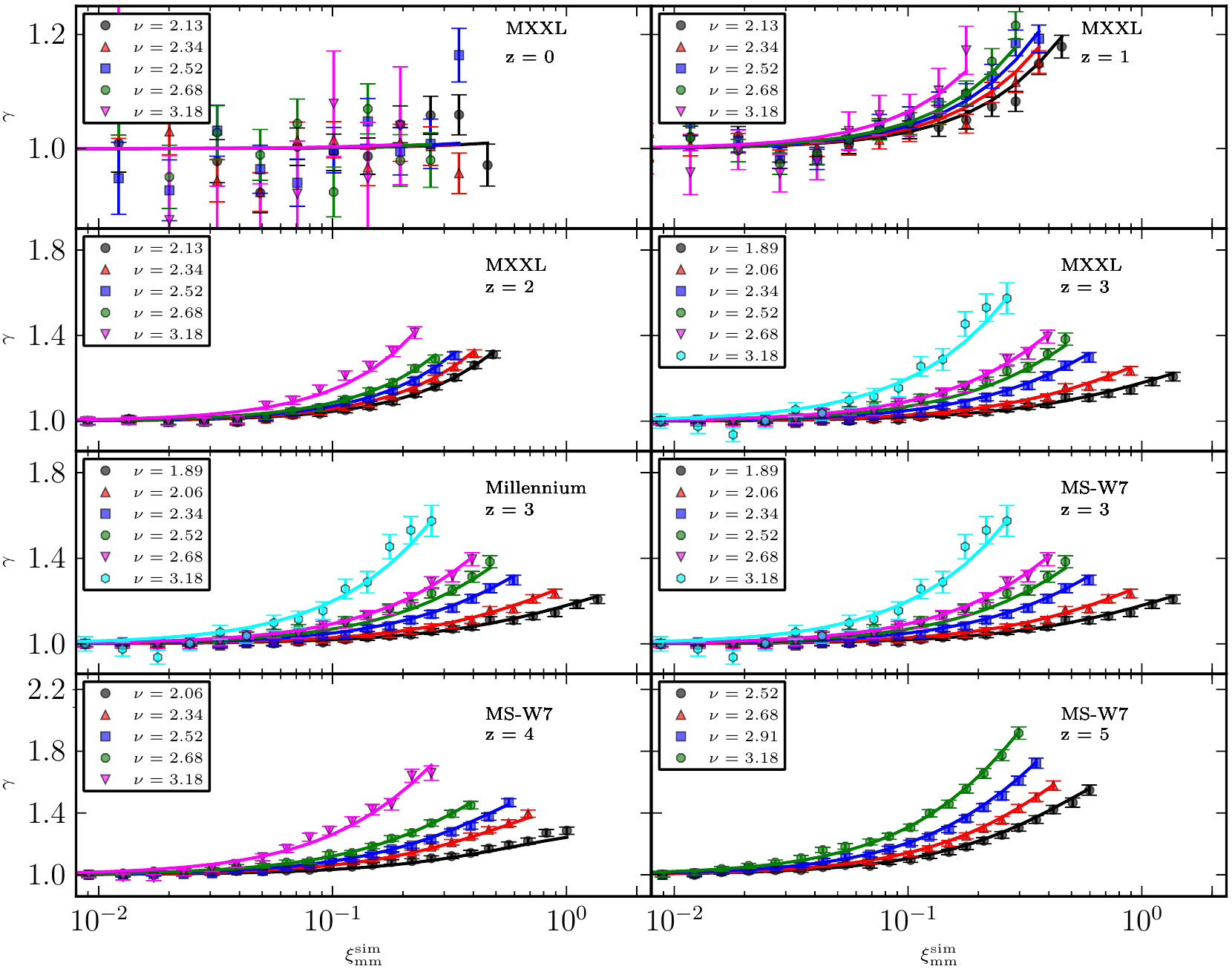} 
\caption[]
{
Our fitting function for the scale dependence of the non-linear bias, 
$\gamma(\xi^{\rm sim}_{\rm mm},\nu, \alpha_m, \Omega_m^z)$, in the redshift range 
$0-5$ as a function of $\xi^{\rm sim}_{\rm mm}(r)$ for various choices 
of $\nu$ (see legend) is shown using solid lines. The N-body simulation 
from which results were measured along with the redshift is labelled on each panel. 
}
\label{fig:gamma_z05}
\end{figure*}

Having obtained the expression for the large scale ($r \geq 10$ Mpc/h) bias 
of rare halos in the redshift range $0-5$, we now wish to calibrate the scale 
dependence $\gamma(M,r,z)$ of $b(r,M,z)$ on quasi-linear scales. 
We first concentrate on the expression for $\gamma(M,r,z)$ of halos in the redshift 
range $2-5$. This is because, as we shall see later, $\gamma(M,r,z)$ probably has an 
explicit dependence on the effective matter density $\Omega_m(z)$ of the 
universe as a function of redshift. We expect to separate out this dependence by focussing 
on high redshifts ($z \geq 2$) where $\Omega_m(z) \approx 1$. 

It is clear from Fig.~\ref{fig:br_universal_fn} that the scale dependent, non-linear bias 
of halos (from the MS-W7 simulation) of a given $\nu(M,z)$ as function of $\xi^{\rm sim}_{\rm mm}(r,z)$ 
and at $z=3,4$ and $5$, agree fairly well with each other. 
In fact, in this case, the agreement between estimates of $b(r,M,z)$ at different 
redshifts is better than 10 \% on quasi-linear scales ($r \leq 15$ $h^{-1}$ Mpc).  
However, Fig.~\ref{fig:br_universal_fn} shows that 
the scale dependence $b(r,M,z)$ of MXXL halos for the 
same $\nu(M,z)$ at $z=2$ and $3$ (which is also expressed as function of 
$\xi^{\rm sim}_{\rm mm}(r,z)$) is quite different from that of MS-W7 halos at higher redshift. 
Thus, $\gamma(M,r,z)$, which accounts for the scale dependence of 
$b(r,M,z)$, cannot be described as a function of just two variables, 
$\xi^{\rm sim}_{\rm mm}(r,z)$ and $\nu(M,z)$. This suggests that the non-linear bias on 
quasi-linear scales is not a simple function of the dark matter power spectrum 
and any fitting function should be a function of other parameters. 

Such an explicit dependence of the halo bias on parameters other than 
the dark matter power spectrum has been discussed in several analytical 
studies \citep{Matsubara1999, blanton_99, 
iliev_scannapieco_03,sheth_tormen_04, gao_2005,Jeong_09, mcdonald_09, 
 lazeyras_16}. 
These studies point out that one may potentially require an infinite 
number of parameters to express the scale-dependent bias.  
On the other hand, most of the available fits to the results of N-body simulations present 
the scale-dependent halo bias as a universal function of the 
dark matter power spectrum \citep{hamana_01,diaferio_03,cen_dong_04,
tinker_05,gao_white_05,reed_09, desjacques_09}. 
We will now investiagate whether the scale-dependence of the halo bias measured 
from the simulations can be expressed as a function of additional 
parameters along with $\xi^{\rm sim}_{\rm mm}(r,z)$ and $\nu(M,z)$.

We find that adding one more parameter can account for all the simulation results in 
the redshift range $2-5$. 
That is, $\gamma(M,r,z)$ at $2 \leq z \leq 5$ can be expressed, 
to sufficient accuracy, as function of three 
variables, $\nu(M,z)$, $\xi^{\rm sim}_{\rm mm}(r,z)$ and $\alpha_m(z)$, an effective power law 
index of $\sigma(M,z)$. This effective power law index is 
defined as
\be
\alpha_m(z) = \df{\log(1.686)}{\log[M_{nl}(z)/M_{col}(z)]}, 
\label{eqn:alpha}
\ee
where the non-linear mass scale, $M_{nl}(z)$, and the collapse 
mass scale, $M_{col}(z)$, at any redshift are masses at which the peak 
heights are, respectively, 1.686 and 1. 
The parameter $\alpha_m(z)$ can be thought of as an effective 
power law index of $\sigma(M,z)$ in the mass range from the collapse to 
the non-linear mass scale (see Appendix A for more details). 
The dependence of $\gamma(M,r,z)$ on $\alpha_m(z)$ can be 
tentatively understood from Fig.~\ref{fig:mnltomcoll}, where we have plotted this ratio 
as a function of $z$. The blue triangles at $z=3,4$ and $5$ are obtained for the MS-W7 
and the red circles at lower redshifts are for the MXXL cosmological parameters. 
The figure clearly shows that $\alpha_m(z)$ is nearly constant in the redshift 
range $3-5$ for the MS-W7 cosmology. However, at $z=2$ where the MXXL 
cosmology is used, $\alpha_m(z)$ is larger.  
Such a difference is perhaps related to the departure from the 
universal nature of non-linear bias as a function of $\nu(M,z)$ and $\xi^{\rm sim}_{\rm mm}(r,z)$.  
Motivated by this, we further investigated whether $\gamma(M,r,z)$ can be expressed as 
a function of $\nu(M,z)$, $\xi^{\rm sim}_{\rm mm}(r,z)$ and $\alpha_m(z)$ and we find that, it is 
indeed possible to obtain a good fit for high-$\sigma$ halos ($\nu > 1)$. 
The resulting fitting function is given by 
\begin{align}
& \gamma(\xi^{\rm sim}_{\rm mm},\nu, \alpha_m) =  \nonumber \\ 
&~~~~~~~~~~~ \left(1 + K_0(1+k_3/ \alpha_m)  \log\left(1+{\xi^{\rm sim}_{\rm mm}}^{k_1}\right) \nu^{k_2}\right) \times \nonumber \\
&~~~~~~~~~~~ \left(1 + L_0(1+l_3/ \alpha_m)  \log\left(1+{\xi^{\rm sim}_{\rm mm}}^{l_1}\right) \nu^{l_2} \right)  
\label{eq:b_r} 
\end{align}
The free parameters are estimated to be $K_0 = -0.0697, k_1 = 1.1682, k_2 = 4.7577, 
k_3 = -0.1561, L_0 = 5.1447, l_1 = 1.4023, l_2 = 0.5823$ and $l_3 = -0.1030$. 
In fitting Eq.~(\ref{eq:b_r}), we have used the non-linear bias estimated 
from simulations for all $\nu(M,z)$ bins in redshift range $z = 2-5$ given 
in Table \ref{tab1}. 
For fitting $\gamma(\xi^{\rm sim}_{\rm mm},\nu, \alpha_m)$, we considered halo correlation 
functions only on scales larger than twice the virial radius of the most massive halo in 
the sample. Moreover, we have restricted our analysis to 
$r \leq 30~h^{-1}$ Mpc.

It is also important to note from Table \ref{tab1} that the masses of 
the halos used in our analysis range typically from $5 \times 10^{10}$ to 
$5 \times 10^{13} h^{-1} M_\odot$. Thus, the high-$\sigma$ halos of 
interest are those expected to host galaxies in the redshift 
range $2-5$.

The expression for $\gamma(\xi^{\rm sim}_{\rm mm},\nu)$ in Eq.~(\ref{eq:b_r}) is 
plotted (solid line) 
in Fig.~\ref{fig:gamma_z25} at different redshifts as a function of 
$\xi^{\rm sim}_{\rm mm}(r)$ at that redshift and for different values of $\nu(M)$. 
In Fig.~\ref{fig:gamma_z25}, the results shown at $z=2$ are from the 
MXXL simulation and those at other redshifts are from the MS-W7 simulation. 
From Fig.~\ref{fig:gamma_z25}, it is clear that the parameterization of 
$\gamma(\xi^{\rm sim}_{\rm mm},\nu, \sigma_{\rm eff})$ using Eq.~(\ref{eq:b_r}) fits the whole 
range of data measured directly from the simulations very well. 
In particular, we note that our fit is consistent with the results from 
simulations to within an overall accuracy of 4\%. 
This suggests that, it is indeed possible to find a fitting function for 
the scale dependence of the non-linear bias in the redshift range $2-5$, 
through $\xi^{\rm sim}_{\rm mm}$, $\nu$ and $\alpha_{m}$. 

In Fig.~\ref{fig:gamma_z25}, we have also plotted in dotted lines, the fitting 
function for $\gamma$ given by \cite{reed_09}. These authors parameterized the 
scale dependence as a function of the large scale bias $b(\nu)$ and 
$\sigma(r,z)$ as
\be
\gamma(b(\nu), \sigma(r,z)) = [1+ 0.03 b^3(\nu) \sigma^2(r,z)]. 
\ee 
It is clear from Fig.~\ref{fig:gamma_z25} that the 
\cite{reed_09} fit compares reasonably well with results from the  
MXXL and MS-W7 simulations at $z=3,4$ and $5$, especially at $z=5$. However, 
their formula is not quite consistent with the MXXL simulation results at $z=2$.     
This is expected,  since the \cite{reed_09} expression for non-linear bias of a 
halo of mass $M$ at a scale $r$  depends only on the dark matter power spectrum 
through the {\it rms} linear density fluctuations on the mass scale $M$ and 
length scale $r$. As noted before, such a simple dependence cannot 
accurately account for the scale dependence of the non-linear bias seen from 
simulations. 

\subsection{The evolution of $\gamma(M,r,z)$ to low redshifts}

Having obtained a fitting function for $\gamma(r,M,z)$ for high-z halos 
in the entire redshift range $2-5$, we now include low-z data to probe  
the evolution of the scale-dependence of non-linear bias from $z=0$ to $5$. 
We note from Table \ref{tab1} that, at $z=0$ and $1$, 
the masses of the rare dark matter halos used in our study ranges from  $10^{12}$ to 
$10^{15} h^{-1} M_\odot$; correspondingly they host galaxies as well as groups and 
clusters. 

In Fig.~\ref{fig:gamma_z05}, we show $\gamma(r,M,z)$ estimated from the simulations 
(symbols) over the full redshift range. The data at $z=3,4$ and $5$ are 
measured from MS-W7 and those at $z=0,1$ and $2$ are from the MXXL simulation.  
It is clear from the figure that at lower redshifts ($z=0$ and $1$) the scale 
dependence of the non-linear bias is rather weaker compared to other redshifts. 
Such a weak scale-dependence of halo bias on quasi-linear scales at lower 
redshifts ($z=1$) can be found also in the the analytic work 
of \cite{scannapieco_barkana_02}.
In particular, at $z=0$, the halo bias increases by only $\sim 10\% $ on 
quasi-linear scales even for the most massive and hence rarest halos 
at that redshift. 

It turns out that one can obtain a fit for the non-linear bias which 
extends to redshift $0$ by adding an additional parameter, $\Omega_m(z)$, 
the matter density of the universe at a given redshift. 
\be
\Omega_m(z) = \df{\Omega_m (1+z)^3}{\Omega_m (1+z)^3 + \Omega_\Lambda}. 
\ee
Thus, the evolution of $\gamma(r, M,z)$ in the redshift range $0-5$ can be 
expressed as a function of four variables, 
$\Omega_m(z)$, $\nu(M,z)$, $\xi^{\rm sim}_{\rm mm}(r,z)$ and $\alpha_m(z)$. 
In particular, we obtained a fitting function for $\gamma(r,M,z)$  
using the non-linear bias estimated from simulations for halos in bins of 
$\nu(M,z)$ in the redshift range $z = 0-5$ given in Table \ref{tab1}.   
As before, we have used 
the correlation functions only on scales larger than twice the virial radius of the 
biggest halo in the sample and smaller than $30$ Mpc/h for the analysis. 
The resulting fitting function is given by, 
\begin{align}
& \gamma(\xi^{\rm sim}_{\rm mm},\nu, \alpha_m,\Omega_m(z)) =  \nonumber \\ 
&~~~~ \left(1 + K_0(1+k_3/\alpha_m) \left(\Omega_m(z)\right)^{k_4} \log\left(1+{\xi^{\rm sim}_{\rm mm}}^{k_1}\right) \nu^{k_2}\right) \times \nonumber \\
&~~~~ \left(1 + L_0(1+l_3/\alpha_m) \left(\Omega_m(z)\right)^{l_4} \log\left(1+{\xi^{\rm sim}_{\rm mm}}^{l_1}\right) \nu^{l_2} \right)  
\label{eq:b_r_full} 
\end{align}
Here $K_0 = 0.1699, k_1 = 1.194, k_2 = 4.311, k_3 =−0.0348, k_4 = 17.8283, L_0 = 2.9138, 
l_1 = 1.3502, l_2 = 1.9733, l_3 =-0.1029$ and $l_4 = 3.1731$. 
The fitting function in Eq.~(\ref{eq:b_r_full}) is plotted as solid lines 
in  Fig.~\ref{fig:gamma_z05} along with data points measured from 
simulations. 
Our fit is in remarkable agreement with data from all the simulations over  
the entire range of redshifts from $0$ to $5$, peak heights and length 
scales. The overall agreement of this fit with the data from the  
simulations is found to be better than 4\%.

\subsection{Halo clustering as a function of the linear matter correlation function}
We have, so far, presented a model for the non-linear 
clustering of dark matter halos as a function of the non-linear 
dark matter correlation function, $\xi^{\rm sim}_{\rm mm}(r,z)$, measured 
from the simulations. 
In this section, we model halo clustering as a function 
of the linear matter correlation function, $\xi^{\rm lin}_{\rm mm}(r,z)$. 
This is well motivated because $\xi^{\rm lin}_{\rm mm}(r,z)$ is 
easier to compute without uncertainties, compared to 
the non-linear matter correlation function. Thus, for all practical 
purposes, it will be convenient to express the non-linear bias as 
a function of $\xi^{\rm lin}_{\rm mm}(r,z)$. 
The linear matter correlation function is computed from the linear 
matter power spectrum $P^{lin}(k,z)$ as  
\be
\xi^{\rm lin}_{\rm mm}(r,z)  = \int\limits_0^\infty \df{dk}{2\pi^2}~k^2~P^{lin}(k,z)~\df{\sin(kr)}{kr}.
\label{eqn:ximmlin} 
\ee

In order to model the non-linear halo bias as a 
function of $\xi^{\rm lin}_{\rm mm}(r,z)$, we first 
define $b_{nl}(r,M,z)$ at any given scale as 
\be
b_{nl}(r,M,z) = \sqrt{\df{\xi^{\rm sim}_{\rm hh}(r,z)}{\xi^{\rm lin}_{\rm mm}(r,z)}}.  
\label{eq:brlin}
\ee
The new definition of $b_{nl}(r,M,z)$ is similar to that given 
by Eq.~(\ref{eq:br}), but uses $\xi^{\rm lin}_{\rm mm}(r,z)$ 
instead of $\xi^{\rm sim}_{\rm mm}(r,z)$. 
Following section~\ref{sec:nlbias} we then express the non-linear 
bias as the product of the scale-independent large scale bias 
$b(\nu)$ and the scale-dependent 
function $\gamma(r,M,z)$ (see Eq.~(\ref{eq:gammadef})).  
A new fitting function is obtained for the large scale bias 
by refitting the free parameters of Eq.~(\ref{eq:tbias}) to 
the large scale bias measured from the simulations. 
The new best fit parameters are given by $A =1.0, a= 0.223, B=1.156, 
b= 2.167, C=-0.748$ and $c=2.167$. 

As before, we find that, the scale-dependence of halo bias $\gamma(r,M,z)$  
can be expressed as a function of $\nu$, $\alpha_m,\Omega_m(z)$ 
and the linear matter correlation function, $\xi^{\rm lin}_{\rm mm}$. 
The fitting function for $\gamma$ is assumed to have the same functional 
form as in Eq.~(\ref{eq:b_r_full}) with 
$\xi^{\rm sim}_{\rm mm}$ being replaced by $\xi^{\rm lin}_{\rm mm}$. 
The free parameters of Eq.~(\ref{eq:b_r_full}) are then determined 
by fitting this equation to the data measured from all the 
simulations in the redshift range $0-5$. 
The new best fit parameters of Eq.~(\ref{eq:b_r_full}) are given 
by $K_0 = 0.000529, k_1 = 1.0686, k_2 = 3.4158, k_3 =-204.1715, 
k_4 = 26.9453, L_0 = 0.448, l_1 = 2.128, l_2 = 3.0222, l_3 =0.226$ 
and $l_4 = 1.691$. 
We emphasize that the new fit for $\gamma$ as function of 
$\xi^{\rm lin}_{\rm mm}$ agrees very well with the simulation data. 
The overall agreement of this fit with the data from 
all the simulations given in Table \ref{tab1} is found to 
be better than 5\%.  

\section{Discussion and conclusions}
We have revisited the problem of modelling the non-linear clustering of rare dark 
matter halos, that collapse from $1-3 \sigma$ fluctuations, on quasi-linear scales.  
In particular, we found using high-resolution N-body simulations that 
the non-linear bias of high redshift galactic 
dark matter halos is strongly scale dependent on 
scales $ \sim 0.5 -10$ $h^{-1}$ Mpc. 
These scales, commonly referred to as quasi-linear scales, correspond to 
scales larger than the typical virial radii of dark matter halos. 
Even though we primarily 
focussed on the clustering of dark matter halos in the redshift range 
$0-5$, our results are applicable to higher redshifts, including the 
cosmic dark ages before the epoch of reionization. 

First, we estimated the correlation functions of dark matter halos 
at $z =2,3,4$ and $5$ from the N-body simulations, in mass bins in the mass range 
$ 10^{11} - 4 \times 10^{12} M_\odot$. 
These are the typical masses of dark matter halos that host LBGs 
and LAEs in the same redshift range and correspond to rarer objects 
collapsing from high $\sigma$ fluctuations \citep{charles_12_clustering}. 
We then showed that, on quasi-linear scales, there is a strong 
discrepancy between the halo correlation functions computed using the scale 
independent, linear halo bias and those measured 
directly from simulations.  
This suggests that the linear bias approximation is not sufficient to explain 
the clustering of high-z, rarer dark matter halos on quasi-linear scales. 

To quantify the non-linear bias of dark matter halos in this redshift range, 
we measured the correlation functions of halos, from simulations, in 
bins of halo peak height, $\nu(M,z)$. The non-linear bias 
is defined as the square root of the ratio of halo and dark matter correlation 
functions (see Eq.~(\ref{eq:br})). 
We found that the non-linear bias of a halo can be expressed as the product 
of the usual scale independent large scale bias $b(M,z)$ and 
a scale dependent function $\gamma(r,M,z)$ (Eq.~(\ref{eq:gammadef})). 
We also obtained a fitting function for $b(M,z)$ which depends 
only on the peak height, $\nu(M,z)$, of dark matter halos. 
This fit compares very well with other formulae for large scale bias 
in the literature \citep{sheth_tormen_99,tinker_10}, especially 
for halos with $\nu \lesssim 2$, collapsing from low $\sigma$ fluctuations. 
For rarer halos with larger values of $\nu$, we obtained a slightly lower value 
for the large scale bias compared to the formula given by \cite{tinker_10}. 
However, as noted before, this could be due to the difference 
between the SO and FOF(0.2) halo finder algorithms respectively used 
by \cite{tinker_10} and in our work. Further, both studies 
use distinct simulations with different cosmological 
parameters for calibrating the bias. 

We find that, for $z=2-5$, the scale dependence of the non-linear bias, 
$\gamma(r,M,z)$, for halos of mass $M$ at any length scale $r$ depends on three 
parameters, the peak height, $\nu(M,z)$, of halos at mass $M$, the dark matter correlation 
function ($\xi^{\rm sim}_{\rm mm}(r,z)$) at that length scale and $\alpha_m$, 
an effective power law index of $\sigma(M)$ at that redshift. 
We obtained a fitting function that describes the scale 
dependence of $\gamma(r,M,z)$ as a function of these parameters 
in the same redshift range. Our fit agrees with the simulation 
results within an accuracy of 4\%.  

The scale dependence of non-linear bias at a scale $r$ is usually 
parametrized in real space using 
$\xi^{\rm sim}_{\rm mm}(r,z)$ \citep{tinker_05} or 
the {\it rms} linear overdensity in uniformly overdense 
spheres of radius $r$, $\sigma(r,z)$ \citep{hamana_01,diaferio_03,reed_09}. 
Both $\sigma(r,z)$ and $\xi^{\rm sim}_{\rm mm}(r,z)$ can be expressed as 
functions of the dark matter power spectrum. 
However, we find that the scale dependence of the bias, as quantified in terms 
of $\gamma(r,M,z)$, is not described by such parametrizations, 
but rather depends on the quantity $\alpha_m(z)$. 
But, to compute $\alpha_m(z)$, one requires only the linear 
dark matter power 
spectrum. Therefore, it can be argued that at high redshifts ($z \geq 2$) 
the non-linear bias is a universal function of the 
dark matter power spectrum. 

We extended our analysis by probing the non-linear 
bias of low redshift, rarer dark matter halos on 
quasi-linear scales, using  MXXL halos at $z=0$ and $1$. Interestingly at   
lower redshifts, especially at $z\sim 0$, the scale dependence of non-linear 
bias is weaker than at high redshifts and 
is within 10-20 \% of the large scale bias measured from 
simulations at any scale. 
We propose a fitting function for the non-linear bias as a function of 
the matter density of the universe at a given redshift ($\Omega_m(z)$) 
along with $\nu(M,z)$,  $\xi^{\rm sim}_{\rm mm}(r,z)$ and $\alpha_m(z)$. 
Remarkably, this fitting function, calibrated using the MS-W7 and MXXL 
simulations, captures the redshift evolution of non-linear bias for a wide 
range of halo masses and length scales within an overall accuracy of $4 \%$.

The dependence of $\gamma(r,M,z)$ on $\Omega_m(z)$ at low redshifts 
breaks the universality of the non-linear bias with respect to the 
linear matter fluctuation field. Thus the observed large scale bias 
of any galaxy population, which depends only on the dark matter 
power spectrum through $\nu(M,z)$, will not uniquely determine the 
scale dependence of the bias. 
This may provide an opportunity to use 
the scale-dependence of halo bias as a valuable tool to probe 
cosmology, particularly the matter density of the universe. 
 
We have also extended our analysis by expressing the non-linear 
bias as a function of the linear matter correlation 
function, $\xi^{\rm lin}_{\rm mm}(r,z)$. 
Here also the non-linear bias is expressed as the product 
of the scale independent large scale bias, $b(M,z)$,  and 
the scale dependent function, $\gamma(r,M,z)$. 
We first obtained a fitting function for the large scale bias 
as a function of $\nu(M,z)$. 
A fitting function for $\gamma(r,M,z)$ is then obtained as a 
function of linear matter correlation function, 
$\xi^{\rm lin}_{\rm mm}(r,z)$, 
along with $\nu(M,z)$, $\alpha_m(z)$ and $\Omega_m(z)$. 
The new fit for $\gamma$ agrees with the data from the 
simulations within an accuracy better than 5\%. 
We emphasize that this model parameterizes the clustering of dark matter 
halos as a function of $\xi^{\rm lin}_{\rm mm}(r,z)$ instead of 
the non-linear matter correlation function. 
Such a model could be quite useful for practical purposes as 
it is easier to compute $\xi^{\rm lin}_{\rm mm}(r,z)$ analytically 
without uncertainties compared to the non-linear matter correlation 
function. 

In general, the halo bias of high redshift, rare dark matter halos is 
significantly non-linear and scale dependent on quasi-linear scales. 
On the other 
hand, at $z=0$, this scale dependence is quite weak and seems to be 
dependent on the matter density of the universe. 
The non-linear bias is expected to have interesting 
implications on observations of the high redshift universe. 
For example, the halo occupation distribution modelling of LBG 
clustering at high-z ($z \geq 3$) usually assumes a linear 
halo bias \citep{hamana_04,hamana_06,hildebrandt_09_acf,lee_09,charles_12_clustering}. 
However, at these redshifts, the typical LBGs collapse from 
$2-3 \sigma$ fluctuations.  Hence, one has to incorporate the non-linear 
bias to improve the the clustering predictions of LBGs on quasi-linear 
scales. 
Thus the non-linear bias could change the predicted shape of the 
two point correlations functions of high redshift LBGs and also LAEs, 
quasars and even the redshifted 21 cm signals from the pre-reionization. 
It would be interesting to explore the implications of the non-linear 
and scale-dependent bias in the high-z universe. 
To do this, one may need to incorporate the effects of 
baryons on the clustering of galaxies through the physics of galaxy 
formation and also of assembly bias. 
This is left for future work.


\section*{Acknowledgments}
CJ thanks Kandaswamy Subramanian, Raghunathan Srianand and Aseem Paranjpaye 
for useful discussions. 
CJ acknowledges partial support from the Institute for Computational 
Cosmology (ICC), Durham University while visiting the ICC and 
Carlos Frenk at the ICC for warm hospitality. 
This work was supported by the Science and Technology Facilities Council 
[ST/L00075X/1]. This work used the DiRAC Data Centric System at Durham 
University, operated by the ICC on behalf of the STFC DiRAC HPC 
Facility (www.dirac.ac.uk). This equipment was funded 
by BIS National E-infrastructure capital grant ST/K00042X/1, STFC capital 
grant ST/H008519/1, and STFC DiRAC Operations grant ST/K003267/1 and 
Durham University. DiRAC is part of the UK's National E-Infrastructure. 

\def\aj{AJ}%
\def\actaa{Acta Astron.}%
\def\araa{ARA\&A}%
\def\apj{ApJ}%
\def\apjl{ApJ}%
\def\apjs{ApJS}%
\def\ao{Appl.~Opt.}%
\def\apss{Ap\&SS}%
\def\aap{A\&A}%
\def\aapr{A\&A~Rev.}%
\def\aaps{A\&AS}%
\def\azh{AZh}%
\def\baas{BAAS}%
\def\bac{Bull. astr. Inst. Czechosl.}%
\def\caa{Chinese Astron. Astrophys.}%
\def\cjaa{Chinese J. Astron. Astrophys.}%
\def\icarus{Icarus}%
\def\jcap{J. Cosmology Astropart. Phys.}%
\def\jrasc{JRASC}%
\def\mnras{MNRAS}%
\def\memras{MmRAS}%
\def\na{New A}%
\def\nar{New A Rev.}%
\def\pasa{PASA}%
\def\pra{Phys.~Rev.~A}%
\def\prb{Phys.~Rev.~B}%
\def\prc{Phys.~Rev.~C}%
\def\prd{Phys.~Rev.~D}%
\def\pre{Phys.~Rev.~E}%
\def\prl{Phys.~Rev.~Lett.}%
\def\pasp{PASP}%
\def\pasj{PASJ}%
\def\qjras{QJRAS}
\def\rmxaa{Rev. Mexicana Astron. Astrofis.}%
\def\skytel{S\&T}%
\def\solphys{Sol.~Phys.}%
\def\sovast{Soviet~Ast.}%
\def\ssr{Space~Sci.~Rev.}%
\def\zap{ZAp}%
\def\nat{Nature}%
\def\iaucirc{IAU~Circ.}%
\def\aplett{Astrophys.~Lett.}%
\def\apspr{Astrophys.~Space~Phys.~Res.}%
\def\bain{Bull.~Astron.~Inst.~Netherlands}%
\def\fcp{Fund.~Cosmic~Phys.}%
\def\gca{Geochim.~Cosmochim.~Acta}%
\def\grl{Geophys.~Res.~Lett.}%
\def\jcp{J.~Chem.~Phys.}%
\def\jgr{J.~Geophys.~Res.}%
\def\jqsrt{J.~Quant.~Spec.~Radiat.~Transf.}%
\def\memsai{Mem.~Soc.~Astron.~Italiana}%
\def\nphysa{Nucl.~Phys.~A}%
\def\physrep{Phys.~Rep.}%
\def\physscr{Phys.~Scr}%
\def\planss{Planet.~Space~Sci.}%
\def\procspie{Proc.~SPIED}%
\let\astap=\aap
\let\apjlett=\apjl
\let\apjsupp=\apjs
\let\applopt=\ao

\bibliographystyle{mn2e}	
\bibliography{ref.bib,sfr.bib}		

\appendix
\section{The effective power law index of $\sigma(M)$ }
For any mass scale $M$, the variance of smoothed density 
contrast $\sigma^2(M) \propto  k^3_M P(k_M) \sigma^2_8 
k_M^{3+n_{\rm eff}}$ \citep{peebles_80}. Here $n_{\rm eff}$ is the effective spectral 
index, which is $\sim -2$ on galactic scales and $-1$ on 
cluster scales. We also have $k_M^{-1} \sim M^{1/3}$. 
Thus we get 
\be
\sigma(M) \propto M^{\df{-(3+n_{\rm eff})}{6}} \propto M^{-\alpha}. 
\label{eq:A1}
\ee
Given the non-linear mass, $M_{nl}$, and collapse mass, $M_{col}$, corresponds to  
the mass scales where $\sigma(M)$ is respectively $1$ and $1.686$, it 
is possible to define an effective power law index $\alpha_m$ as 
\be
\df{\sigma(M_{col})}{\sigma(M_{nl})} = 1.686 = \left( \df{M_{col}}{M_{nl}} \right)^{-\alpha_m}
\ee
Thus we have
\be
\alpha_m = \df{\log(1.686)}{\log(M_{nl}/M_{col})} = 0.2269 \left[\log \f{M_{nl}}{M_{col}}\right]^{-1}
\ee

\end{document}